\documentclass[journal]{IEEEtran}
\usepackage[T1]{fontenc}

\usepackage{graphicx}
\usepackage{cite}
\usepackage{picinpar}
\usepackage{amsmath}
\usepackage{url}
\usepackage{flushend}
\usepackage{colortbl}
\usepackage{soul}
\usepackage{multirow}
\usepackage{pifont}
\usepackage{color}
\usepackage{alltt}
\usepackage{enumerate}
\usepackage{siunitx}
\usepackage{epstopdf}
\usepackage{pbox}
\usepackage{threeparttable}

\usepackage{siunitx}
\usepackage[acronyms]{glossaries}
\glstoctrue
\newacronym{3gpp}{3GPP}{3rd Generation Partnership Project}
\newacronym{cdf}{CDF}{Cumulative Distribution Function}
\newacronym{rf}{RF}{Radio Frequency}
\newacronym{dnn}{DNN}{Deep Neural Network}
\newacronym{iq}{IQ}{Inphase and Qudrature}
\newacronym{aut}{AUT}{Antenna Under Test}
\newacronym{fcl}{FC Layer}{Fully Connected Layer}
\newacronym{bnl}{BN Layer}{Batch Normalization Layer}
\newacronym{lte}{LTE}{Long-Term Evolution}
\newacronym{mac}{MAC}{Medium Access Control}
\newacronym{ml}{ML}{Machine Learning}
\newacronym{nn}{NN}{Neural Network}
\newacronym{pdf}{PDF}{Probability Density Function}
\newacronym{ta}{TA}{Timing Advance}
\newacronym{sdr}{SDR}{Software-Defined Radio}
\newacronym{crb}{CRB}{Cramér–Rao bound}
\newacronym{mcrb}{MCRB}{Modified Cramér–Rao bound}
\newacronym{adam}{ADAM}{Adaptive Moment Estimation}
\newacronym{relu}{ReLU}{Rectified Linear Unit}
\newacronym{papr}{PAPR}{Peak-to-Average Power Ratio}
\newacronym{sgd}{SGD}{Stochastic Gradient Descent}
\newacronym{snr}{SNR}{Signal to Noise Ratio}
\newacronym{ota}{OTA}{Over the Air}
\newacronym{pa}{PA}{Power Amplifier}

\makeatletter
\let\oldtheequation\theequation
\renewcommand\tagform@[1]{\maketag@@@{\ignorespaces#1\unskip\@@italiccorr}}
\renewcommand\theequation{(\oldtheequation)}
\makeatother

\usepackage[nameinlink,capitalize]{cleveref}

\ifCLASSOPTIONcompsoc
    \usepackage[caption=false, font=normalsize, labelfont=sf, textfont=sf]{subfig}
\else
\usepackage[caption=false, font=footnotesize]{subfig}
\fi

\definecolor{limegreen}{rgb}{0.2, 0.8, 0.2}
\definecolor{forestgreen}{rgb}{0.13, 0.55, 0.13}
\definecolor{greenhtml}{rgb}{0.0, 0.5, 0.0}

\begin{document}
\title{Robust and Efficient Fault Diagnosis of mm-Wave Active Phased Arrays using Baseband Signal}
\author{
	\vskip 1em
	Martin H. Nielsen,
	Yufeng Zhang, 
	Changbin Xue,
	Jian Ren, \IEEEmembership{Member,~IEEE},
	Yingzeng Yin, \IEEEmembership{Member,~IEEE},\\
	Ming Shen, \IEEEmembership{Senior Member,~IEEE}, and Gert F. Pedersen, \IEEEmembership{Senior Member, IEEE}
	\thanks{
		Manuscript received Month August, 2021; revised Month April, 2022; accepted Month May, 2022.
		
		M. H. Nielsen is with the Department of Electronic Systems, Aalborg University, Aalborg, 9220 Denmark (e-mail: mhni@es.aau.dk).
		
		Y. Zhang is with the National Space Science Center, Chinese Academy of Sciences, Beijing, 100190 China, also with the University of Chinese Academy of Sciences, Beijing, 100049 China, also with the Department of the Electronic Systems, Aalborg University, Aalborg, 9220 Denmark (e-mail: zhangyufeng15@mails.ucas.edu.cn).
		
		C. Xue is with the National Space Science Center, Chinese Academy of Sciences, Beijing, 100190 China (e-mail: xuechangbin@nssc.ac.cn).
		
        J. Ren and Y.Z. Yin are with the National Key Laboratory of Antennas and Microwave Technology, Xidian University, Xi'an, China (e-mail: renjian@xidian.edu.cn; yzyin@mail.xidian.edu.cn).
		
        M. Shen (Corresponding author) and G. F. Pedersen are with the Department of Electronic Systems, Aalborg University, Aalborg, 9220 Denmark (e-mail: mish@es.aau.dk;  gfp@es.aau.dk).
	}
}

\maketitle
\begin{abstract}
One key communication block in 5G and 6G radios is the active phased array (APA). To ensure reliable operation, efficient and timely fault diagnosis of APAs on-site is crucial. To date, fault diagnosis has relied on measurement of frequency domain radiation patterns using costly equipment and multiple strictly controlled measurement probes, which are time-consuming, complex, and therefore infeasible for on-site deployment. This paper proposes a novel method exploiting a Deep Neural Network (DNN) tailored to extract the features hidden in the baseband in-phase and quadrature signals for classifying the different faults. It requires only a single probe in one measurement point for fast and accurate diagnosis of the faulty elements and components in APAs.

Validation of the proposed method is done using a commercial 28 GHz APA. Accuracies of 99\% and 80\% have been demonstrated for single- and multi-element failure detection, respectively. Three different test scenarios are investigated: on-off antenna elements, phase variations, and magnitude attenuation variations. In a low signal to noise ratio of 4 dB, stable fault detection accuracy above 90\% is maintained. This is all achieved with a detection time of milliseconds (e.g 6~ms), showing a high potential for on-site deployment.

\end{abstract}


\section{Introduction}

\begin{figure}[t!] 
\centering
\includegraphics[width=0.9\linewidth]{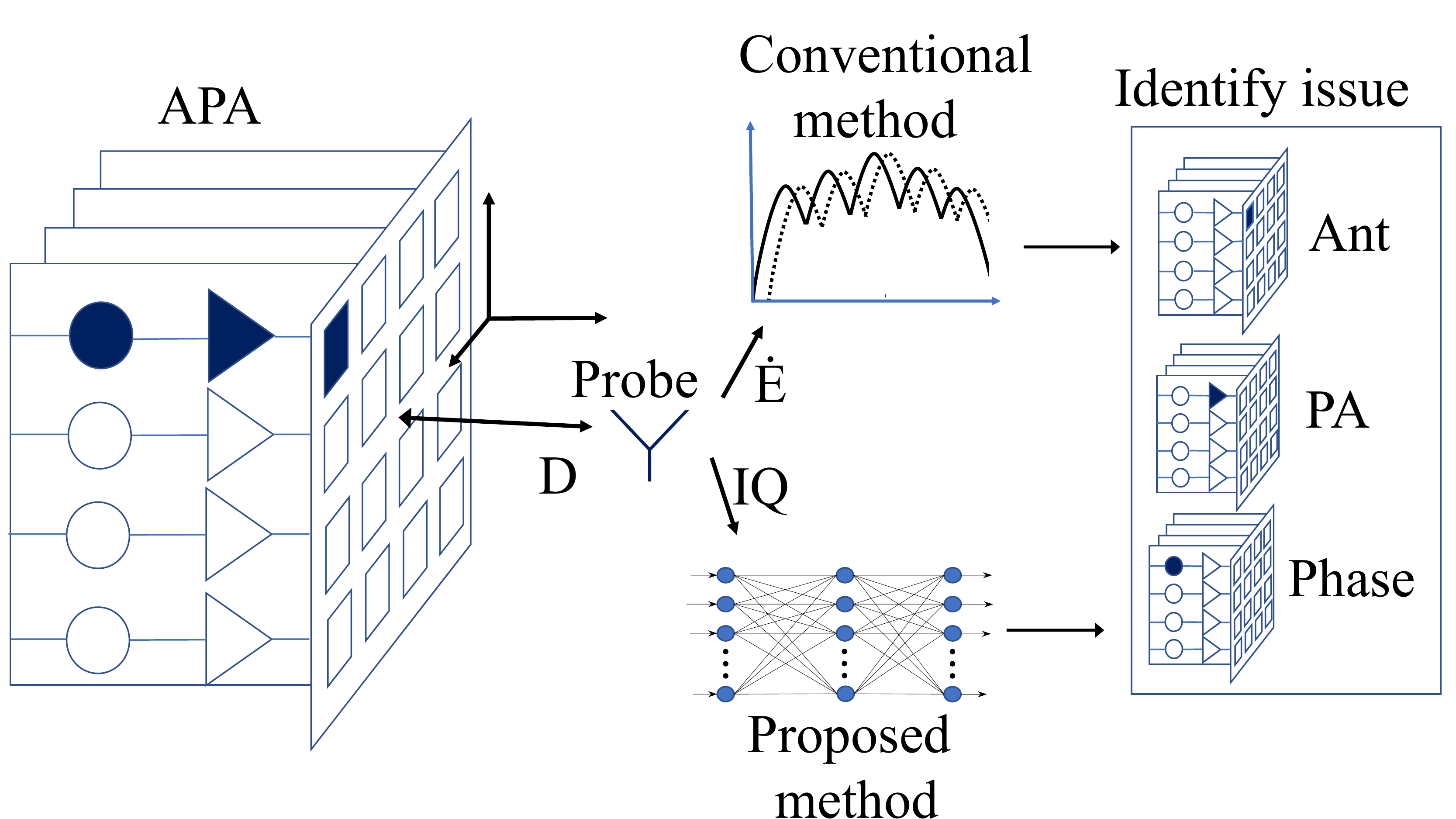}
 \caption{APA diagnosis concept, where the conventional approach is using a math estimator based on measured electric field denoted $\dot{E}$ and antenna pattern, and the novel approach proposed in this paper is using DNN and acquired IQ baseband signal at the receiver. The potential faulty components are highlighted with color. The three potential errors are phase shifters represented using circles, power amplifiers represented using triangles, and antenna elements represented by squares, respectively.}
 \label{fig:convAI-concept}
\end{figure}

\IEEEPARstart{T}{he} commercial availability of 5G communication systems is estimated to triple in the next coming years. The primary enabler of 5G and the next evolution 6G is the use of mmWave active phased arrays (APAs) as the primary communication front end system \cite{boccardi2013disruptive}. 
Quality of Service (QoS) is one of the prime motivators for the evolution of communication systems, fault diagnosis of communications systems has never been more important to meet the QoS user expectation. 
It is critical to identify the failures in the APA efficiently, or even on-site, to make repairs easy and keep the downtime of devices low. Furthermore, failures are not only limited to antenna elements as conventional fault diagnosis techniques have focused on \cite{8739801,8826038,1192,175113,1406243,5752825,4558321,1143531,8443435,8286892,8248776,8765381,Iglesias2008,616650,385367,6236069,6058598,267353}. 
Faults can also happen in the front end circuits such as power amplifiers (PAs) or phase shifters, which makes the fault diagnosis a challenging multi-dimensional problem as illustrated in Fig. \ref{fig:convAI-concept}. 

  \begin{figure*}[ht!]
    \centering
    \includegraphics[width=0.95\linewidth]{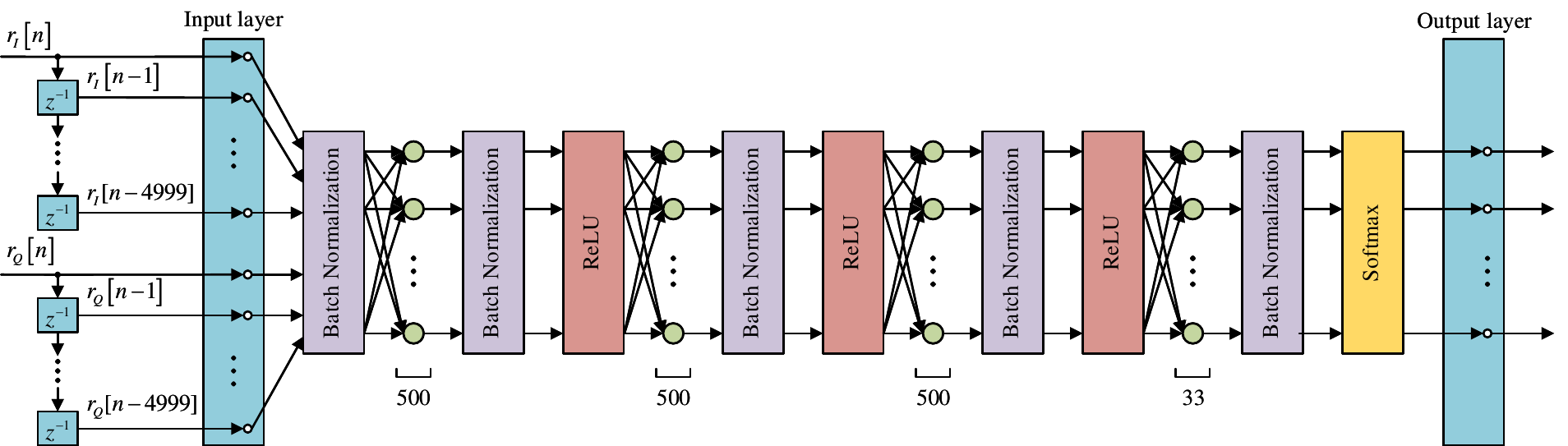}
    \caption{The neural network tailored for APA fault diagnosis is a simple 4-layered feedforward architecture, where each hidden layer has a size of 500 nodes. }
    \label{fig:ArchNN}
\end{figure*}

Fault detection in antenna arrays has been a research topic for many years \cite{8739801,8826038,1192,175113,1406243,5752825,4558321,1143531,8443435,8286892,8248776,8765381,Iglesias2008,616650,385367,6236069,6058598,267353}. 
The techniques for determining faults vary in the mathematic formulation and use different measurement methods. However, they all have in common that they use the electric field component of the antenna array. The measurements also need to be in either near field or far-field and use either 1 probe with a lot of measurements needed or using a plane of probes with fewer measurements needed. 

The most prevalent method used in the literature has been the rotating element electric field vector (REV) method \cite{REVmethod,Fengchun}, hence it is used as the base comparison. 
The special thing with REV is that compared to other approaches for diagnosis of APAs which mostly limits malfunctions to passive antenna elements and phase errors \cite{8739801,8826038,1192,175113,1406243,5752825,4558321,1143531,8443435,8286892,8248776,8765381,Iglesias2008,616650,385367,6236069,6058598,267353,FastOnly,ComplexMethod}. The REV method does not look at passive elements but can detect amplitude, phase, and element faults with very accurate estimations hence why it is also used for calibration arrays before deployment. 

Another fault diagnosis method is the inverse scatter problem (ISP) or more specifically electromagnetic inverse problem (EIP) where the antenna array is seen as the scatter. Instead of relying on the antenna pattern, it uses the time-domain of the measurement to make a tap-delay model of the received signal. This measurement can then be reshaped using mathematical equations to produce an answer for what shape the scatter has. This has been a thoroughly researched topic \cite{8657753,9502427,8476623,8741152,8884635}. However, it needs multiple input sources/measurement probes to be used. Therefore array diagnosis efficiencies of conventional methods have been significantly limited and not suited for on-site deployment.  

\gls{dnn} has shown promises in many fields of fault diagnosis \cite{7076586,7386639,8511076,8114247}. 
Recent research in planar array diagnosis has shown that \gls{dnn} can identify failures if given the full RF domain antenna radiation pattern for each failure type  \cite{Vakula2009, Patnaik, Patnaik_2, 8746510}. 

The multidimensional problem of identifying multiple types of faulty components is tricky and is only solvable using full radiation patterns from anechoic chamber measurements. 5G measurement setups are very sensitive to the positioning of probes due to the scaling down of signal wavelengths at mm-wave frequencies. RF domain-based synthetic method is therefore not suitable for making measurement setup simpler. 
\cite{8739801,8826038,1192,175113,1406243,5752825,4558321,1143531,8443435,8286892,8248776,8765381,Iglesias2008,616650,385367,6236069,6058598,267353}.
Newer research has shown that using a comparable method to REV using DNNs it is possible to reduce calculation complexity. But the methods do not reduce measurement complexity \cite{8746510,9301192,9362206,9581935}. 

This paper proposes a new APA diagnosis technology using baseband IQ signals shown in Fig. \ref{fig:convAI-concept}. 
In-phase and Quadrature (IQ) baseband signals are used in many modern digital systems, and the baseband signal characteristics are unique to the APA. 
Baseband signals typically have a frequency range of up to several hundreds of MHz and are much easier to sample and process. Research has also shown that IQ signals are easier to handle than amplitudes and phases \cite{8795484}. 
Hence, to simplify the measurement setup and data acquisition this uniqueness of the baseband signal is exploited in this paper.

The challenge for conventional synthetic methods to use baseband IQ signals is that it is not a trivial task to model the detailed dependence of the IQ signals on the APA components. According to the theory of deep learning, however, a DNN with sufficient parameters can model any mathematical equations \cite{Ming_book}. The DNN used in this work consists of 4 layers, and 5549147 trainable parameters, providing compelling characterization capability. 
Therefore the DNN can capture the "hidden" features in the IQ signals, which cannot be characterized by conventional techniques.
The novelty of this work is the proposed method's capability to detect not only the failures in the antenna array but also the failures of active components in the front end. This is new and different from our previous work done at sub 3~GHz with planar arrays \cite{telfor}. 
Moreover, the proposed method eliminates the need for multiple measurements at restricted locations opening up new avenues for measurement and test setups, which greatly increases its potential for on-site deployment. To our best knowledge, work that can provide similar outcomes has not been reported yet. 

This paper is organized as follows. Section II describes the new proposed method. In this Section the network, faults, measurement setup and training details are described. In section III the experimental results and test of the model are presented. Section IV is a discussion of the proposed method and a comparison with the conventional methods given earlier. Finally, the paper is concluded in Section V.

 \section{Proposed Method} 
The exploitation of the baseband signal starts with understanding the basic model of the communication system. The communication system can be seen as a dynamic nonlinear system with memory as the basic model description. 
This leads to the following distortions constellation, intersymbol interference and adjacent channel interference. Any small imperfection in the front end system will have impact on these and will be part of the baseband signal. Being a dynamic nonlinear system it can be described by Volterra theory where complex baseband input signal x(n) and output y(n) are related as 
\begin{multline}
 y(n) = \sum_{k=1}^\infty \sum_{n_1} \ldots \sum_{n_k} h_k(n_1,n_2, ... n_k)\\ \prod_{r=1}^p x(n-r_r) \prod_{q=p+1}^k x(n-n_q)^{*}
\end{multline}
where $h_k(n_1,n_2,...,n_k)$ are the Volterra kernels. 
Since any imperfection in the transmitter will have an impact on the output due to the model basis for the Volterra series this can be exploited for the classification and detection of failures in the APA. Using the IQ baseband over a channel that is assumed to be static while transmitting it is possible to create a machine learning algorithm for classification that is better suited for monitoring. 
 \begin{figure*}[ht!]
    \centering
    \includegraphics[width=0.8\linewidth]{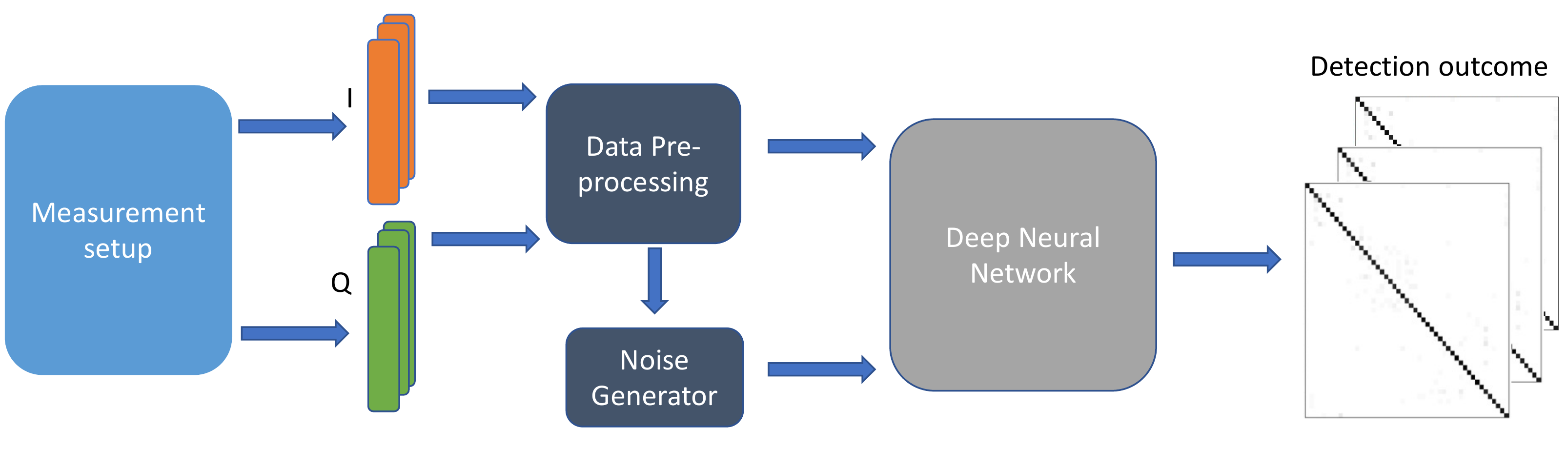}
    \caption{The training process of the DNN, where the IQ signal is acquired from measurement setup and pre-processed for the DNN to handle. Here noise can be added for further robustness tests. After going through the DNN different detection outcomes (e.g. confusion matrix) can be given.}
    \label{fig:IllustrationArchNN}
\end{figure*}

The \gls{dnn} used to identify the different faults is trained using supervised learning to map the different classes to the faults.
DNN has been used instead of more complicated networks like CNN or RNN. These other networks can potentially decrease the number of neurons needed in the network but are more complex to feed and pre-process the data. Since pre-processing and labeling the data is the most costly task it has been decided to keep the network simple.
For simplifying the training each instance of training data is grouped in batches were 200 training data samples are grouped and processed through the network. Going through all the training data available is one epoch. 
The programming language used for the \gls{dnn} is Python and the machine learning library used is PyTorch v 1.4.

\subsection{Architecture of DNN}
The \gls{dnn} is composed of four different types of layers, including \gls{fcl}, \gls{bnl}, activation function layer and hidden layers. The full network architecture can be seen in Fig. \ref{fig:ArchNN}. 
In the \gls{fcl}, weights and biases are expressed by $\mathbf{W}_i$ and $\mathbf{B}_i$ where $i$ denotes the $i$th layer. The output of the FC $i$th layer can be expressed as 
\begin{equation}
    \mathbf{y}_i = \mathbf{W}_i \mathbf{x}_i + \mathbf{B}_i,
\end{equation}
where $x_i$ is the input to the $i$th layer. The number of neurons in the layer is determined by iterative tests to achieve the best performance. 

The \gls{bnl} normalizes the mean and variance of the input data to 0 and 1 respectively and then gives the input data a new mean and variance corresponding to the new dimension of the data. 
This reduces the time cost of the entire training significantly \cite{pmlr-v37-ioffe15}. The output of a \gls{bnl} can be written as 
\begin{equation}
    \hat{y_i} = \gamma \frac{y_i - E[y_i]}{\sqrt{\text{var} [y_i]+\epsilon}+\beta},
\end{equation}
where $\gamma$ and $\beta$ are the scaling factors for the mean and variance in the new dimensional space. Further, $\gamma$ and $\beta$ are learnable parameters. $\epsilon$ is a constant parameter that prevents denominator from being zero. Generally, it is set to $0.001$. 

The output of the \gls{bnl} is fed into the activation function layer which is chosen to be the \gls{relu} function.
The \gls{relu} activation function outputs the positive part of its input and everything else as zero. 
\begin{equation}
    f(x) = x^{+} = \max (0,x),
\end{equation}
where $x$ is the input to a neuron. ReLu is chosen over Leaky Relu, since the IQ data can be represented by the magnitude and phase, and rectifying the sine wave does not have a significant impact on these two key factors and the sine wave are horizontally symmetric. Instead, using half period of the signal reduces the chance for the model to capture too detailed features (e.g. noise) to avoid overfitting. Testing also confirms this since we haven't seen an increase in accuracy due to using leaky ReLu.
This procedure transfers using forwards propagation the raw data from the first layer to the last layer. The useful features of the raw data are then extracted layer by layer. After each batch, the weights in the \gls{dnn} are updated according to the loss which is evaluated by cross-entropy loss in combination with a softmax function
\begin{equation} \label{eq:loss_p}
   \ell(k, \text{\boldmath$\pi$}) = -\ln \left( \frac{ \mathrm{e}^{\pi_k} }{ \sum_i \mathrm{e}^{\pi_i}} \right),
\end{equation}
where $k$ is the index of the target class and \text{\boldmath$\pi$} are the unnormalized posterior class probabilities which are the output of the last layer of the network.
To optimize the network weights and biases, the popular \gls{sgd} is chosen in combination with an adaptive learning rate that decreases by a factor of $\mathrm{e}
^{-1}$ each time the accuracy score of the system stops and stagnates after two epochs. 
This is to ensure that the gradient descent keeps decreasing to a global minimum without introducing overfitting issues in the network. 
The \gls{sgd} optimizer is given as
\begin{equation}
    Q(w) = \frac{1}{n} \displaystyle\sum_{i=1}^{n} Q_i(w),
\end{equation}
where $w$, is the parameter that minimizes $Q(w)$ and is found by solving the following equation

\begin{equation}
    w - \eta \nabla Q(w) = w - \eta \displaystyle\sum_{i=1}^{n}\frac{\nabla Q_i(w) }{n},
\end{equation}
where $\eta$ is the learning rate of the \gls{dnn}. 
The final layer in the \gls{dnn} is the output layer which handles the classes that are to be determined. This layer can change size depending on how many fault scenarios are to be classified and can be expanded to fit new fault scenarios.
 \begin{figure}[htb!]
    \centering
    \includegraphics[width=0.8\linewidth]{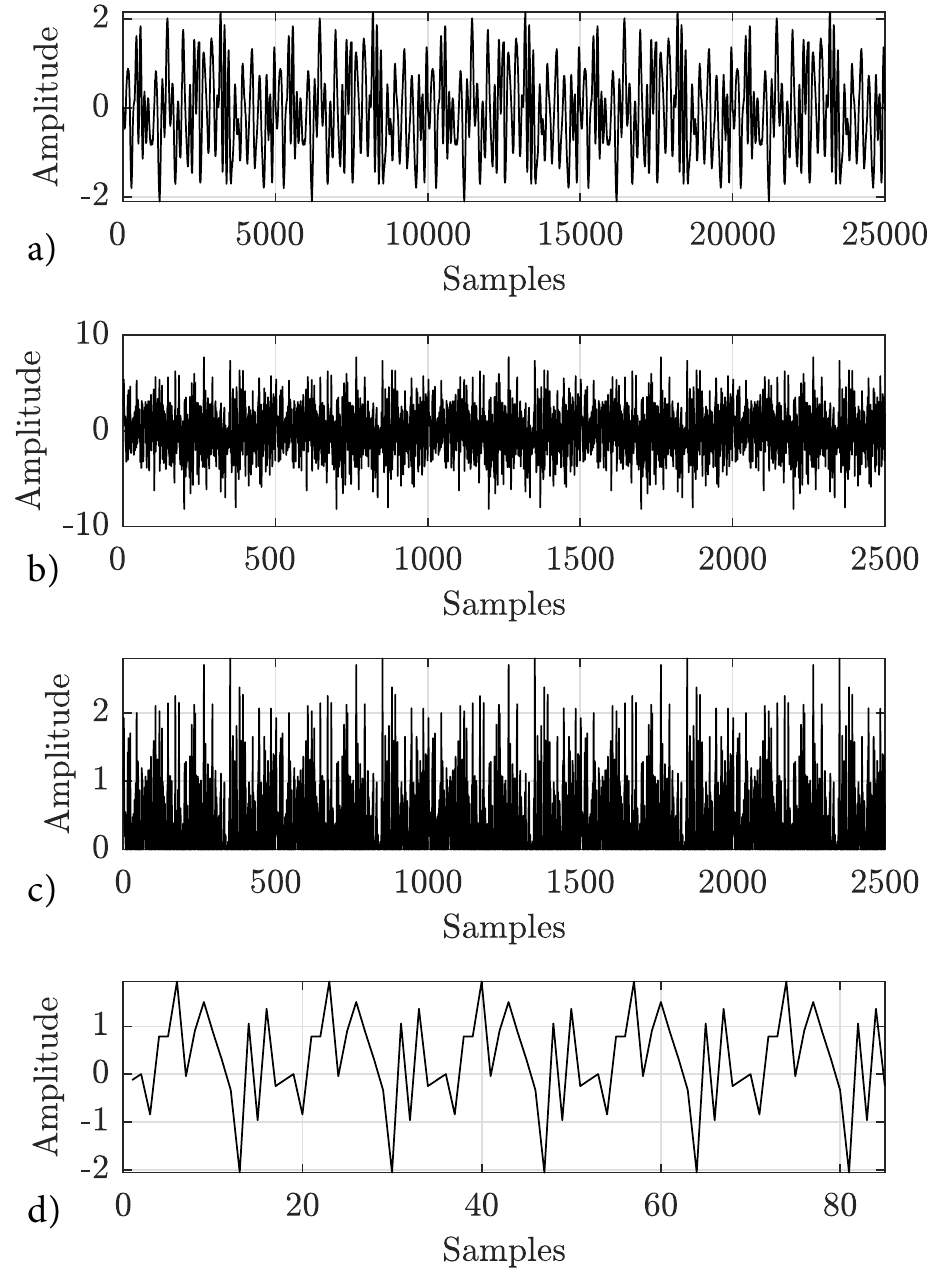}
    \caption{Outputs of different layers showing the influence of each layer of the DNN. a) is the Batch Layer, b) is the Linear layer, c) is the ReLU layer and d) is the output layer.}
    \label{fig:LayerShow}
\end{figure}

\subsection{Fault Scenarios}
The different fault scenarios investigated in this work include three major components used in the APA: antenna element, power amplifier, and phase shifter. 
The different fault scenarios are established by controlling the APA using a software program provided by AMOTECH \cite{amotech}.

\subsubsection{Antenna Element Malfunction} 
Single antenna element malfunction is one antenna element being turned off using AMOTECH software controller while the whole APA is transmitting. Multi-element failure is difficult to show since with 16 elements there are a total of 120 combinations for 2 element failure, for 3 element failure there are 560, for four 1820, etc.
This becomes a very complex measurement issue and too many classes for multi-class classification to solve. Instead, multi-label fault detection can be used. With multi-label classification, the measurement issue persists but the classification issue is no longer growing too large but can be kept to be 16 classes for element detection since multi classes can be predicted. 
If all 16 elements can fail the combination of potential failures is 65535, which is very impractical to collect. 
Instead, this paper limits failures to be grouped into failure categories of faults as 1,2,..,6. Errors of over 6 faulty elements are too many element failures for the power to not have dropped significantly enough that the array is deemed not working. 
The goal of the test is for the DNN to determine the correct antenna element being turned off and distinguish the different IQ waveforms that are being received.

\subsubsection{Magnitude Attenuation Issue}
The goal of this test is to determine if the DNN can detect magnitude imbalance caused by undesired attenuation. In the test, the signal attenuation was controlled by the AMO software with a resolution of 0.5 dB(the lowest resolution available).
The test is carried out by attenuating one signal path with 0.5 dB using the AMO software and then measuring 10 times as outlined in the measurement procedure.

\subsubsection{Phase Fault Detection}
The goal of this test is to determine the DNN's capabilities for handling phase faults. 
Due to limitations of the AMOTECH 0404 phase controller, 1bit phase difference is the lowest phase error possible to introduce while not breaking the APA. 1bit phase difference for the AMOTECH A0404 is 5$^\circ$ of phase shift. 
Each of the 16 antenna elements is then shifted 5$^\circ$ out of phase from the rest one by one to mimic the absolute minimum phase fault that can happen.  

\subsection{Data Collection and Measurement Setup}\label{sec:measurementSetup}
Validation of the DNN was done using a 3 GHz LTE signal, compliant with the 3GPP downlink OFDM modulated also to be used in 5G and satellite high throughput communication \cite{patent:20100068993,6650319}. The signal has a peak to average power ratio of 10.6 dB generated by the R$\&$S SMBV100A  signal generator. 

\begin{figure}[tp!]
    \centering
    \includegraphics[width=0.8\linewidth]{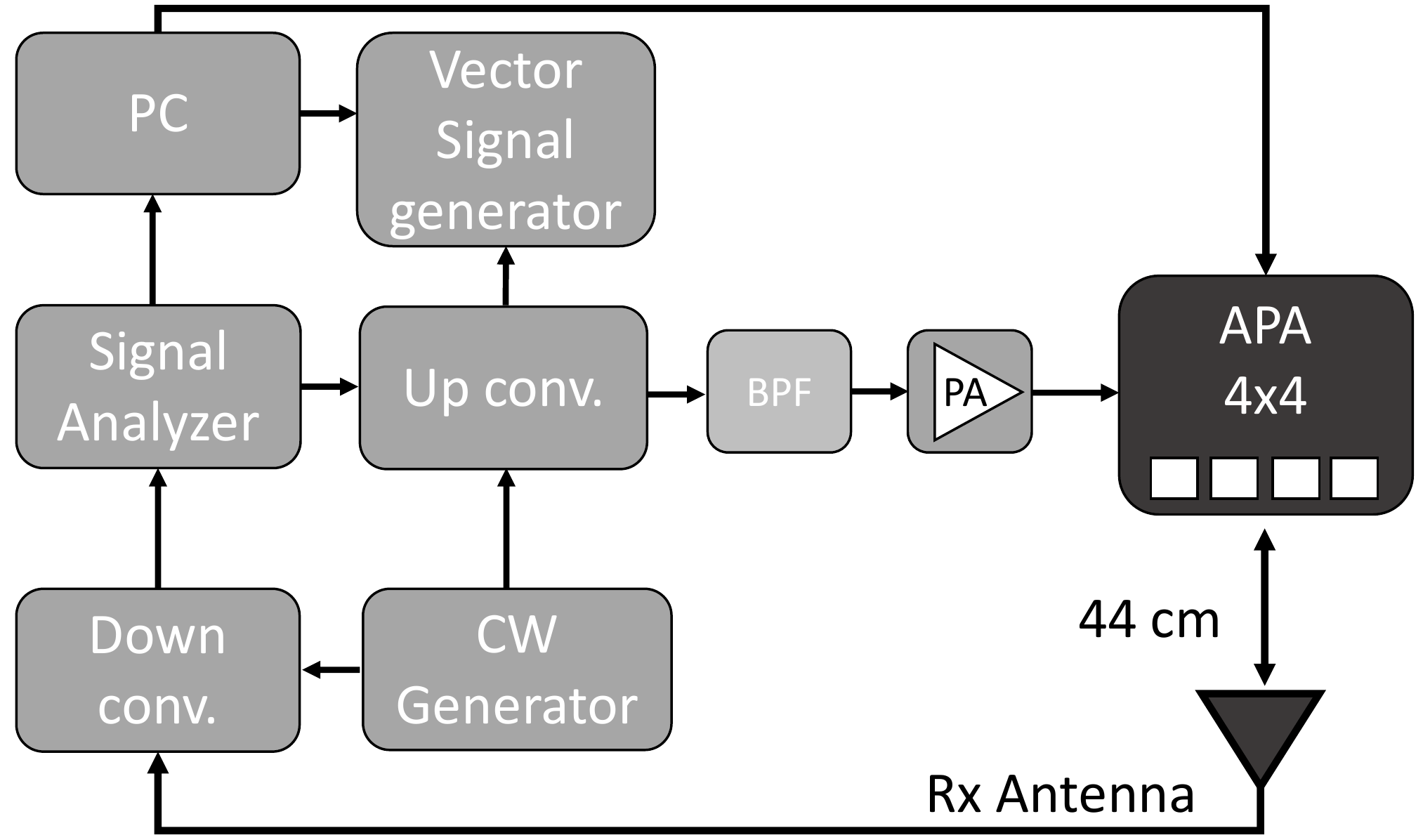}
    \caption{Block diagram of the measurement setup.}
    \label{fig:MeasBLockSetup}
\end{figure}

\begin{figure}[tp!]
    \centering
    \includegraphics[width=0.9\linewidth]{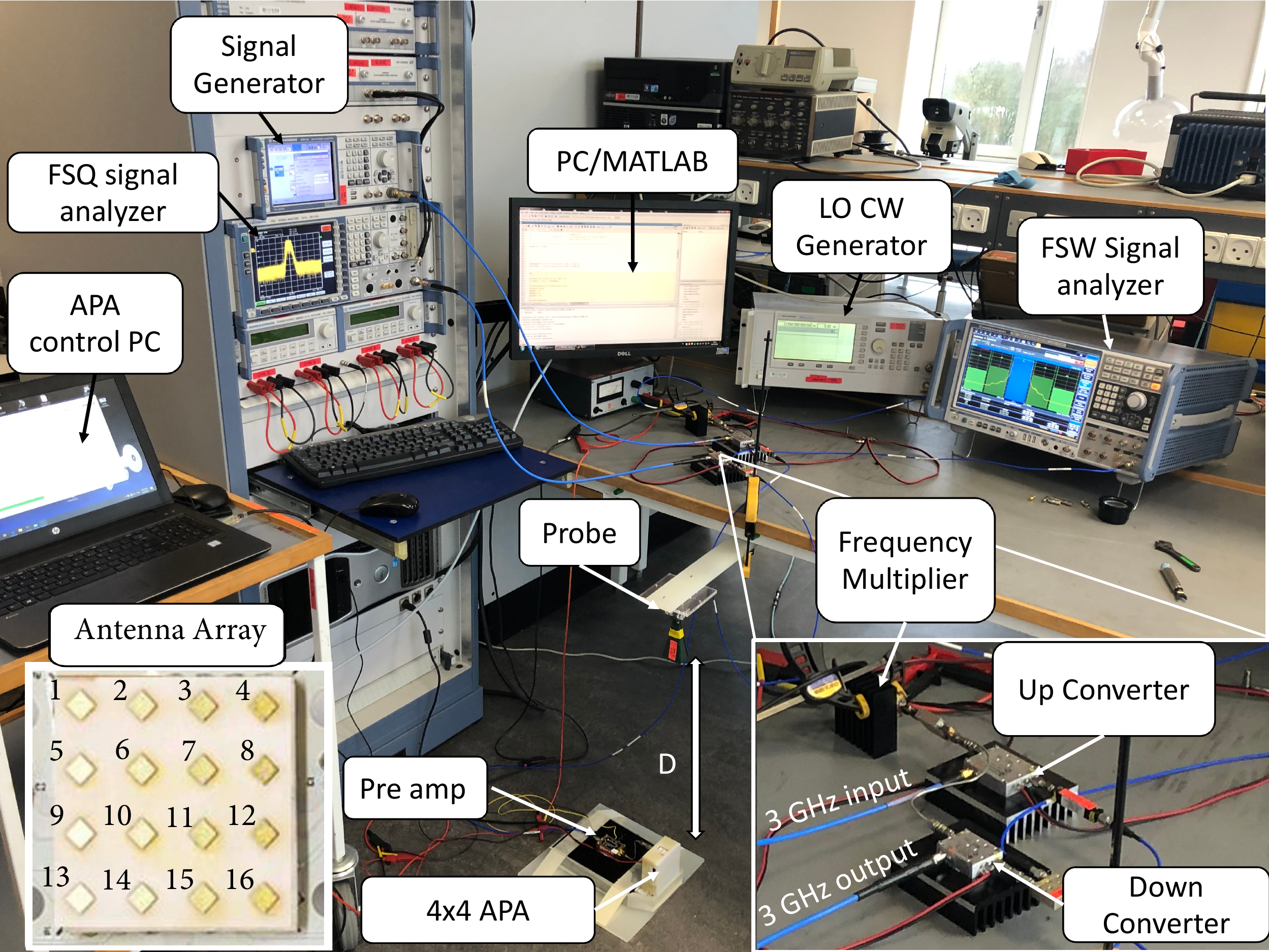}
    \caption{The data acquisition measurement setup.}
    \label{fig:Setup}
\end{figure}

The 3 GHz OFDM signal from the generator is converted to 28 GHz using an up-converter.
A continuous-wave (CW) signal has been multiplied by two into 25 GHz for up-conversion and down-conversion as illustrated in Fig. \ref{fig:MeasBLockSetup}. The leakage from the local oscillator in the up-converted 28 GHz modulated signal is filtered out using a bandpass filter. To avoid any nonlinearity in multiplier and up-converter, the signal levels in these stages are kept in the linear region of these devices according to their specifications. 
The 28 GHz signal is then fed to AMOTECH A0404 \cite{amotech} that includes 4 Anokiwave AWMF-0158  beam forming devices and a 4 by 4 patch antenna array as shown in Fig. \ref{fig:Setup}.
The data is captured by the observation horn antenna placed 42 wavelengths away (44 cm) and aligned with the main beam at 0 deg.

The measurement setup for the 4 by 4 AMO APA is shown in Fig. \ref{fig:Setup}. The captured signal is split into two signal paths. One is analyzed in the spectrum analyzer (R$\&$S FSW 67GHz) for monitoring the actual ACPR. The other is down-converted to a 3 GHz signal and captured by an (R$\&$S FSQ) spectrum analyzer for getting access to I and Q data.
It is chosen to only conduct experiments on single failure detection. This means that only one antenna element can be off, have the wrong phase shift, or radiate less power. 
The signal is captured 10 times with a small random time interval. 
The procedure for taking measurements is as follows: 
\begin{enumerate}
\item I and Q waveform for the LTE signal is uploaded by MATLAB from the PC to the vector signal generator.  
\item The AMO is manipulated to have a fault using custom software to control the different circuits and antennas. 
\item The I and Q  at the receiver are then captured from the signal analyzer 10 times with a random time interval between each measurement.
\item Another antenna element or circuit is chosen to be faulty and the measurement is repeated for all 16 antennas. 
\end{enumerate}

\subsection{Training of the DNN}
\begin{figure}[tp!]
    \centering
    \includegraphics[width=0.9\linewidth]{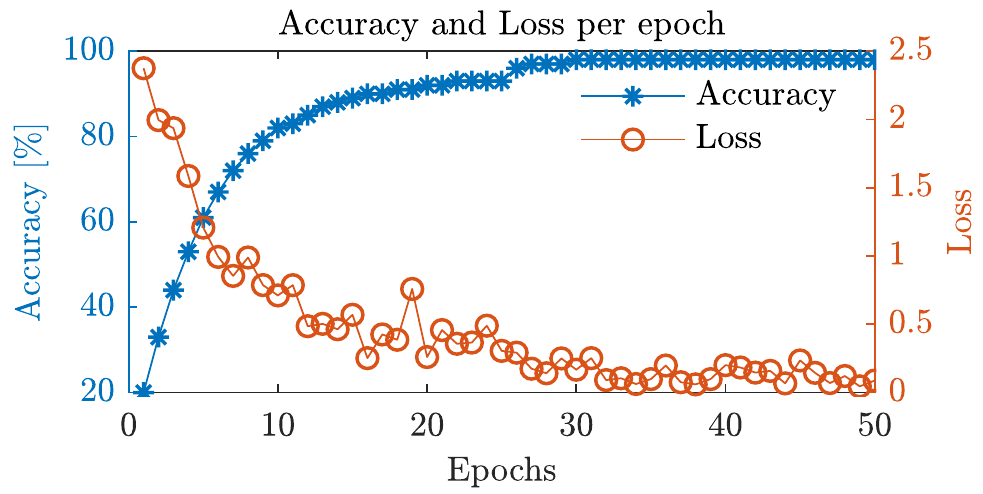}
    \caption{The accuracy and loss versus epochs during training.}
    \label{fig:Acc_Loss_vec_Big}
\end{figure}

For each fault scenario, 10 measurements of $5 \cdot 10^6$ samples at a sample rate of 10~kHz are captured using the spectrum analyzer. 
The 10 measurements are done at random times. This is to ensure the training data is intolerant to time shifts.
\begin{figure}[tp!]
    \centering
    \includegraphics[width=0.9\linewidth]{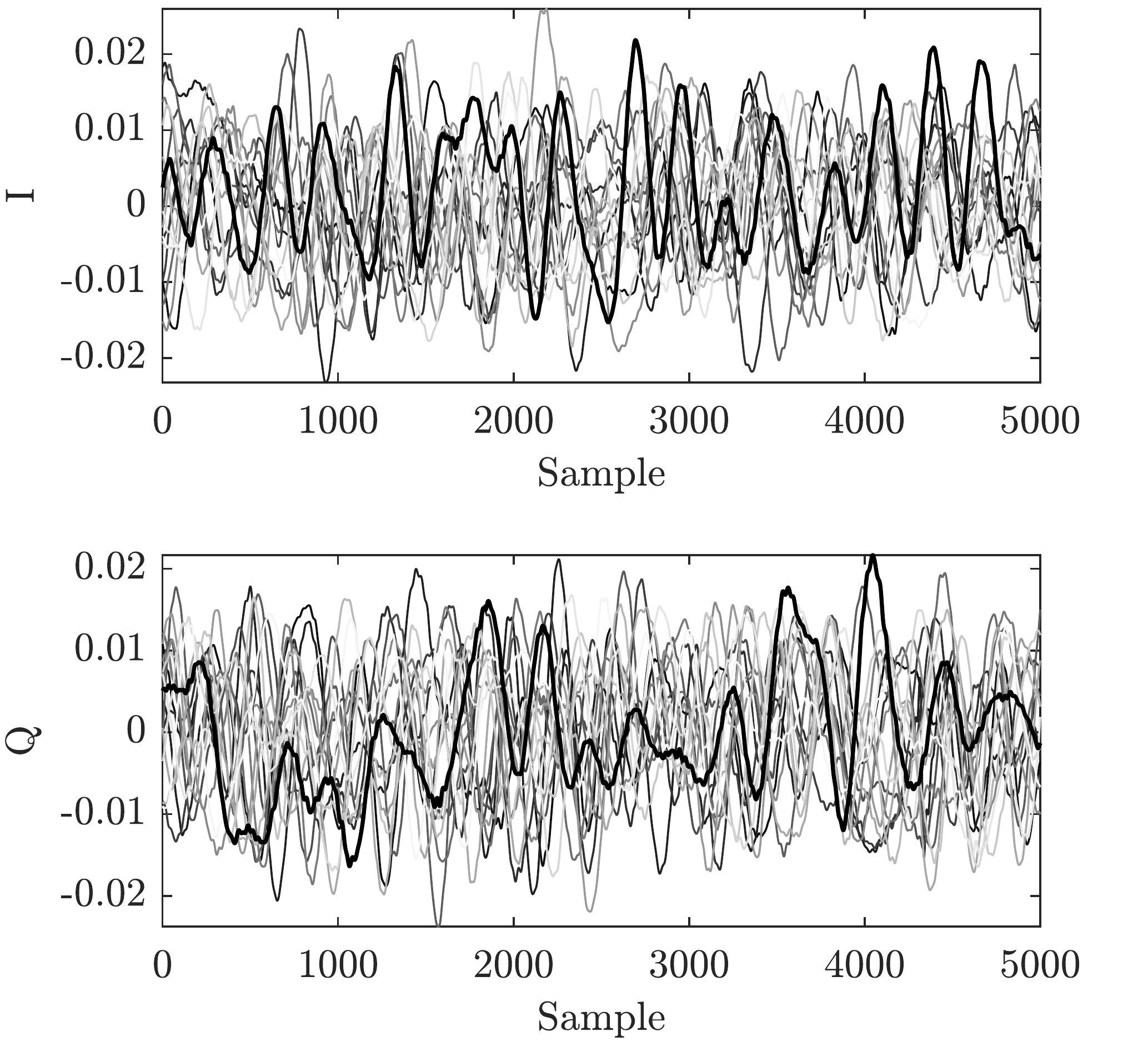}
    \caption{Minibatch of measured IQ signals (two examples highlighted with thicker line width) for each of the 17 different magnitude faults.}
    \label{fig:IQ_plotOf_ATTdata}
\end{figure} 
For training data each class is given as 10 measurements of 5 million samples. To make it easier to handle the data in memory on the GPU it is split into training samples of 5000 I \& 5000 Q samples which are interpreted as one data sample of size 10000 when training the DNN. This expands the number of trainable data samples the DNN can re-iterate over. For testing, a new measurement of 5 million IQ samples per class is done. It is also split into smaller data samples of the same size. 
Fig. \ref{fig:IQ_plotOf_ATTdata} shows different IQ signals features and illustrates that the human eyes or synthetic approaches with close form equations characterized simply characterize the features. 
But as can be seen in the results section, DNN can classify the IQ signals for different faulty cases very well. 
Pytorch does not support complex numbers. Therefore, the input signal is split into the real and imaginary parts of the IQ sequence. The neural network will find a concatenation between the two in training.
The measured IQ data are split randomly into 70\% and 30\% to form training and testing sets, respectively.  
The IQ data is put as the first 5000 samples if I and then the next 5000 samples of Q as one vector of size (1,10000) for the input of the neural network.
The training data is passed through the network over multiple epochs to learn all the features of the network. 
After each epoch, the network gets an accuracy score. To keep improving the DNN and make sure no overfitting happens adaptive learning rate is implemented. If the accuracy score of the network does not improve for three epochs, the learning rate is lowered by a factor of ten.
This is illustrated in Fig. \ref{fig:Acc_Loss_vec_Big}. 

The loss of the network logarithmically decreases while the accuracy exponentially increases as the training data goes through more epochs. 
It is evident that even though the cross-entropy plateaus in the last training steps, the accuracy keeps improving epoch by epoch until it nearly reaches 100 \%.
The DNN has a  final accuracy of 99 \% and a loss very close to 0. 

\section{Experimental Validation}\label{sec:results}
\begin{figure}[tp!]
    \centering
    \includegraphics[width=0.85\linewidth]{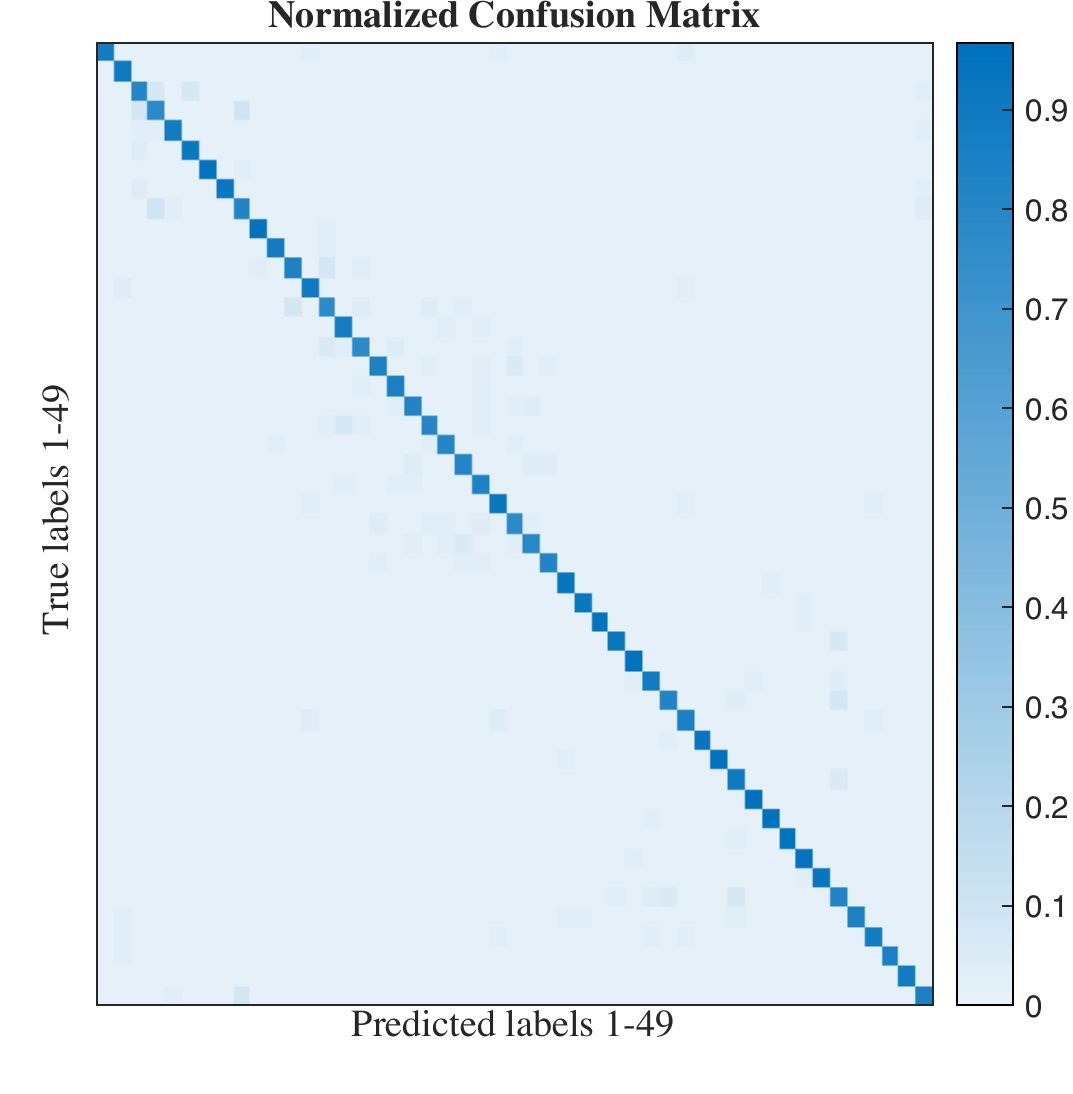}
    \caption{Normalized confusion matrix showing all 50 fault testing cases.}
    \label{fig:Conf_matrix_BIG}
\end{figure}

To evaluate how well the DNN performs in terms of accuracy, the test data is passed through the trained DNN. The results are represented in a confusion matrix shown in Fig. \ref{fig:Conf_matrix_BIG}. 
From the test data, the accuracy of predicting the correct class is between 88 \% and 99 \% dependent on which class is predicted. These are very good results considering the DNN has to determine which of the 49 fault classes it is. Hence, the DNN will with a minimum of 88 \% confidence correctly identify the failure in the AMO APA. 
The different labels in Fig. \ref{fig:Conf_matrix_BIG} are denoted by 1 for no faults, by 2-17 for antenna off, by 17-33 for attenuation with 0.5 dB, and by 34-49 for a phase shift of 5 degrees. 
Each failure is then distinguishable as a label for the DNN to output and the failure label can then be mapped to the component malfunction. 
In Fig. \ref{fig:Conf_matrix_BIG} the resulting accuracy of each class can be seen. The difference between leakyReLu and ReLu has been investigated together with a comparison to existing methods which can be seen in Tab. \ref{tab:comparison}.
The confusion matrix shows a diagonal matrix where the prediction is between 90\% and 98\% meaning the DNN can correctly recognize the true class from the test data. Any wrong prediction will be represented as a shade that is not on the diagonal. Further testing has been done for multi-element testing, here faults are grouped in clusters for easier prediction, there are 1-3 element failures, full-chip failure meaning a full chip has failed and 2 PA and 2 phase shifter faults. To predict every failure in the test data set takes 1.1~s while it takes 0.006~s for a single measurement to be predicted on an NVIDIA TITAN RTX GPU. This makes the on-site deployment of the proposed method promising. And this is especially true as the needed I and Q signals are easy to acquire using a simple receiver front end.

\begin{figure}[tp!]
    \centering
    \includegraphics[width=1.0\linewidth]{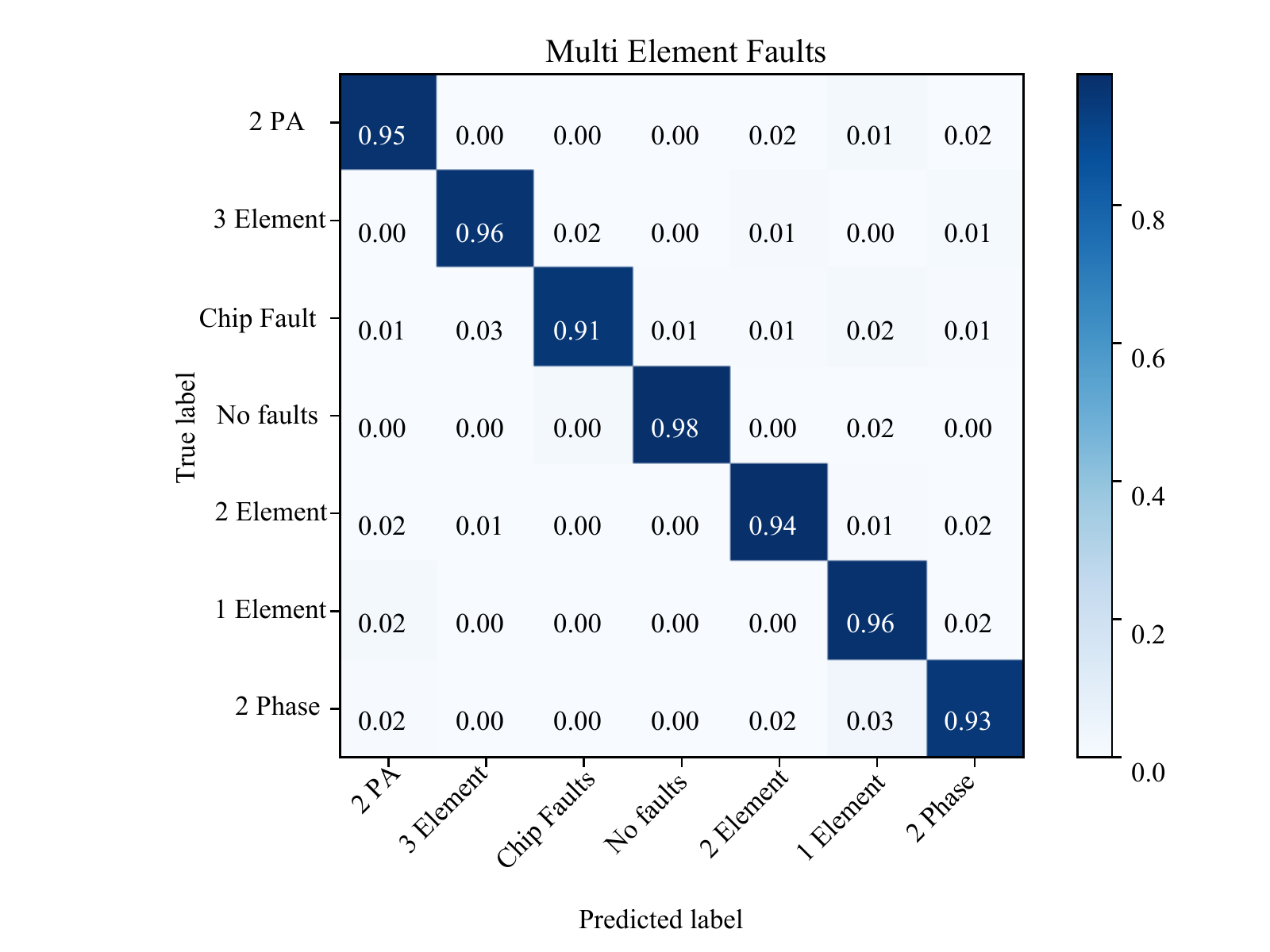}
    \caption{{Normalized confusion matrix for multi-element fault detection for up to 6 faults.} }
    \label{fig:Conf_matrix_multi_element}
\end{figure}
The resulting test is shown in Fig. \ref{fig:Conf_matrix_multi_element}, here it is possible to show the clear separation between different fault cases similar to the performance shown for single element detection.

\subsection{Noise's Effect on Diagnosis}
\begin{figure}[htb!]
    \centering
    \includegraphics[width=0.9\linewidth]{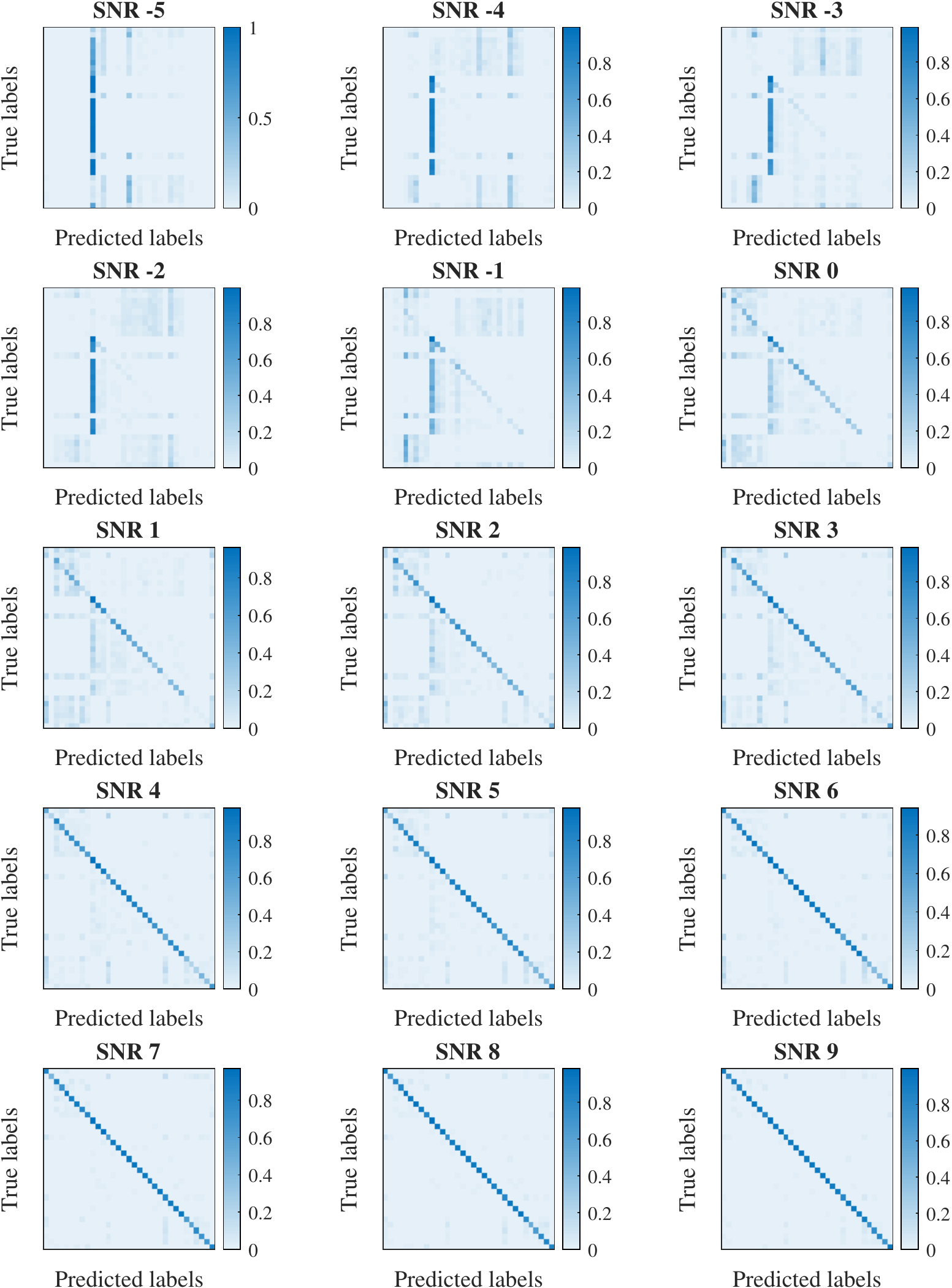}
    \caption{Multiple confusion matrixes with SNR -5-9 dB.}
    \label{fig:MultiConfusion}
\end{figure}

To determine how robust the DNN is, a Signal to Noise Ratio (SNR) test is done.  
To demonstrate this, the test data is polluted with Average white Gaussian noise (AWGN), with a Signal to Noise Ratio (SNR) ranging from -5 dB to 9 dB. 
The resulting performance of the DNN is displayed in Fig. \ref{fig:MultiConfusion}. 

In Fig. \ref{fig:MultiConfusion}, when SNR is  (-5 - -2) one class is predicted for all cases(black straight line down). 
Fig. \ref{fig:MultiConfusion} shows that when the SNR level is between -1 to 3, the addition of noise does not impede the prediction, and from SNR levels of above 4 dB, the DNN model predicts consistently the correct class though with less accuracy.

\begin{figure*}[tp!]
    \centering
    \includegraphics[width=0.95\linewidth]{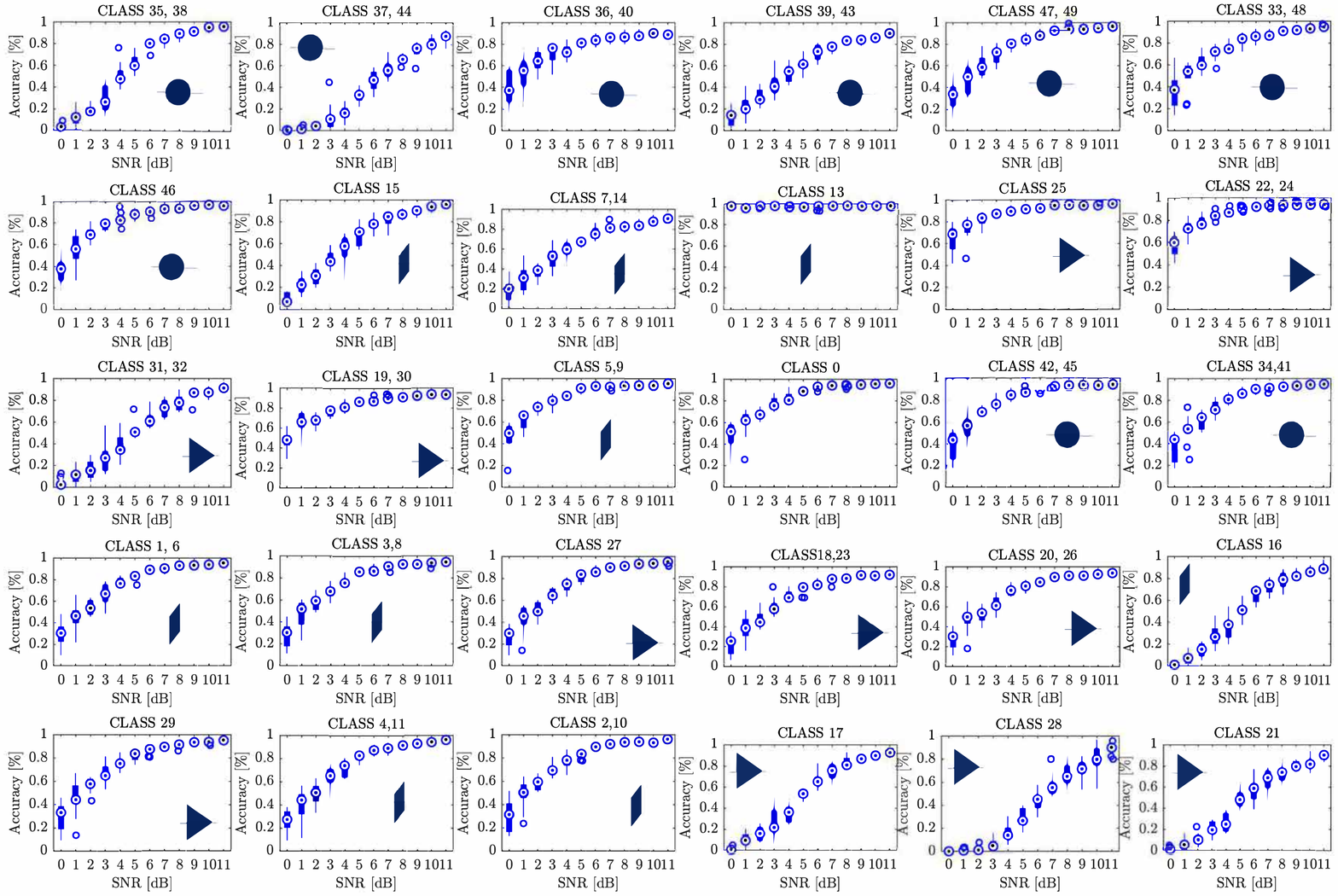}
    \caption{Statistical accuracy versus SNR box-plots for three typs of failures: antenna element, PA, and phase shifter. Class 0 is no failures, classes 1-17 are antenna on off detection, 18-33 is amplifier faults and 34-49 is phase shifter faults. Since some classes have same distribution of detection accuracy they have been plotted together to reduce space. }\label{fig:Boxplot1}
\end{figure*}

\subsection{Statistical Analysis of Performance in SNR}
To test the Statistical performance of the DNN different distributions of the 70/30 split data are used.  It is possible to see how well the DNN is trained and what worst-case and best-case scenarios for data are. By varying the random split of data 10 times, the batches are different from each other with enough variation to test the network's robustness to variations in test data. During this test, all three types of failures, antenna elements, PAs, and phase shifters, have been included. 
Box plots of accuracy versus SNR are used to illustrate the difference of runs over SNR as shown in Fig. \ref{fig:Boxplot1} to illustrate the statistical variance over 10 runs. 
By further looking into the distributions of the different SNR values, it is very clear how the distribution of predictions behaves in different SNR values. For high SNR values, the predictions cluster around 80-100 \% while for low SNR values the distribution gets spread out across the x-axis. 


\begin{table}[]
\caption{Comparison with Previous Works.\label{tab:comparison}}
\centering
\begin{threeparttable}
\begin{tabular}{ccccc}
\hline
Reference & Method  & Accuracy & \begin{tabular}[c]{@{}c@{}}Robustness \\ Estimate\end{tabular} \\ \hline
 \cite{REVmethod} & REV & 100.0\% & 0\%  \\
 \cite{FastOnly} & Fast only & 90.0\% & N/A  \\
 \cite{ComplexMethod} & Complex signal & 100\% & N/A  \\
 \cite{Fengchun} & OTA calibration & 100\% & 50\% \\
\cite{Vakula2009} & DNN &  80.0\% & Yes  \\
 \multirow{2}{*}{\textbf{This work} }& \textbf{DNN} &\textbf{99.9\%} & \textbf{90\%} \\
 & \textbf{DNN LeakyRelu} & \textbf{93.8\%} & \textbf{86.7\%} \\ \hline
\end{tabular}
\end{threeparttable} 
\end{table}

\section{Discussion}
The DNN can distinguish the 48 faulty classes and 1 not faulty class from each other with an accuracy of between 90 \% and 99 \% while the faults are of the smallest possible deviation to be introduced in the commercial APA of 0.5 dB magnitude in a single PA and 5$^\circ$ phase shift per antenna element. 
The DNN's biggest advantage compared to radiation pattern approaches is the ability to use the baseband signal single measurement to predict the faulty components' location and failure type. 
Conventional and other deep learning approaches can guarantee the correct prediction of antenna elements however it requires multiple antenna probes or measurements points. 
Big measurement setups are not needed after training as the model only requires simple IQ baseband signals. Further, since the trained DNN model only takes up 22.3 MB of space, it can be used in many devices with limited storage space.

Multi-element classification is shown to be possible, with the difficulty of determining which element is faulty is was not possible to be solved with the method proposed in this paper alone. Multi-label classification is a possible solution to help solve this, however it comes with significant challenges. 
Simply using IQ data is of too small dimension to directly use multi-label classification. The dimensionality of the data makes multilabel classification only work as a coin toss and is therefore not presented in this paper. Further investigation is needed to solve this problem with more advanced machine learning techniques and measurement setup.

EIP/ISP are related problems but also very different from the proposed solution.  This work uses 5G OFDM baseband signal of IQ at a normal sample size to determine faults in a transmitter at a normal frame size.EIP/ISP needs tap delay measurements and this requires a different measurement setup/different pre-processing of the data to get the needed results. This works baseband approach is a new approach to the same problem and could be beneficial in using for over the air (OTA) in situ fault diagnosis.

The DNN is adaptable for different deployment scenarios using new data and transfers learning schemes to update the DNN quickly. This can make it possible to expand the methods for larger antenna arrays. 

Instead of having the APA work sub-optimal or not at all, a simple reconfiguration of the APA steering vector can be done to compensate for malfunctioning hardware as long as errors are kept under 2 faulty elements, otherwise, the power drop is significant for this small 16 element array. 
Moreover, the DNN shows that even at SNR above 4 dB the accuracy of the predictions is maintained above 90 \%. However, at SNR below 4 dB, the DNN has difficulties and one class ( class 13)  is predicted with above 98 \%. However, it always gets predicted even when SNR is below 4 dB. 
It can therefore be concluded that it is possible to trust the DNN when SNR is above 4 dB but when below 4 dB the results should not be given 100 \% trust. 
A reason for this behavior is because the model has found a strong weight for class 13, and when noise is polluting the IQ signal so much that the signal is no longer recognizable from what the DNN has previously encountered this class 13 will be predicted. 
To potentially eliminate this and be able to handle much larger noise in the signal the DNN should be trained with high noise signals and is future work for this method. 
But it is very impressive to achieve the robustness of the model without noise addition during training. More details have been provided, but out of range is a bit more difficult. As known, a deep learning classifier does not understand out of range and it will be trying to put it into the learned classes. Hence it has not been included as it is seen as arbitrary that out of range does not work for deep learning systems which is one of the drawbacks of the proposed system.

Other possible future work for this method includes the distance relationship between probe position for training data and testing data, the impact multiple failures will have on the detection accuracy of the array, and the impact of noise in testing and training data. 

\section{Conclusion} \label{sec:conclusion}
This paper presents a new deep learning-based method for fault diagnosis of active phased arrays (APAs) widely used in 5G and LEO satellite communication systems.     
The main contribution of this work is the development of the DNN for multiple fault detection while eliminating the need for a big number of measurement points, time-consuming procedures, and costly chambers paving the way for on-site deployment of the fault diagnosis technology. 
The proposed fault diagnosis technique was tested both for accuracy and robustness using a 28 GHz commercial APA with 48 experimental fault scenarios. The trained deep neural network can distinguish component failures between single path magnitude attenuation in the gain down to 0.5 dB, phase variation of 5$^\circ$, and array element failure. Prediction accuracy of up to 99 \% for scenarios without the added noise and above 90\% in presence of significant noises (e.g. SNR of 6 dB) has been achieved. The concept proposed in this work has the potential to be extended to on-site fault diagnosis for more system blocks (e.g. switches, filters, etc.) in advanced communication systems such as 5G and 6G communication systems.

\bibliographystyle{IEEEtran}
\bibliography{IEEEexample}

\vspace{-1.0cm}
\begin{IEEEbiography}[{\includegraphics[width=1in,height=1.25in,clip,keepaspectratio]{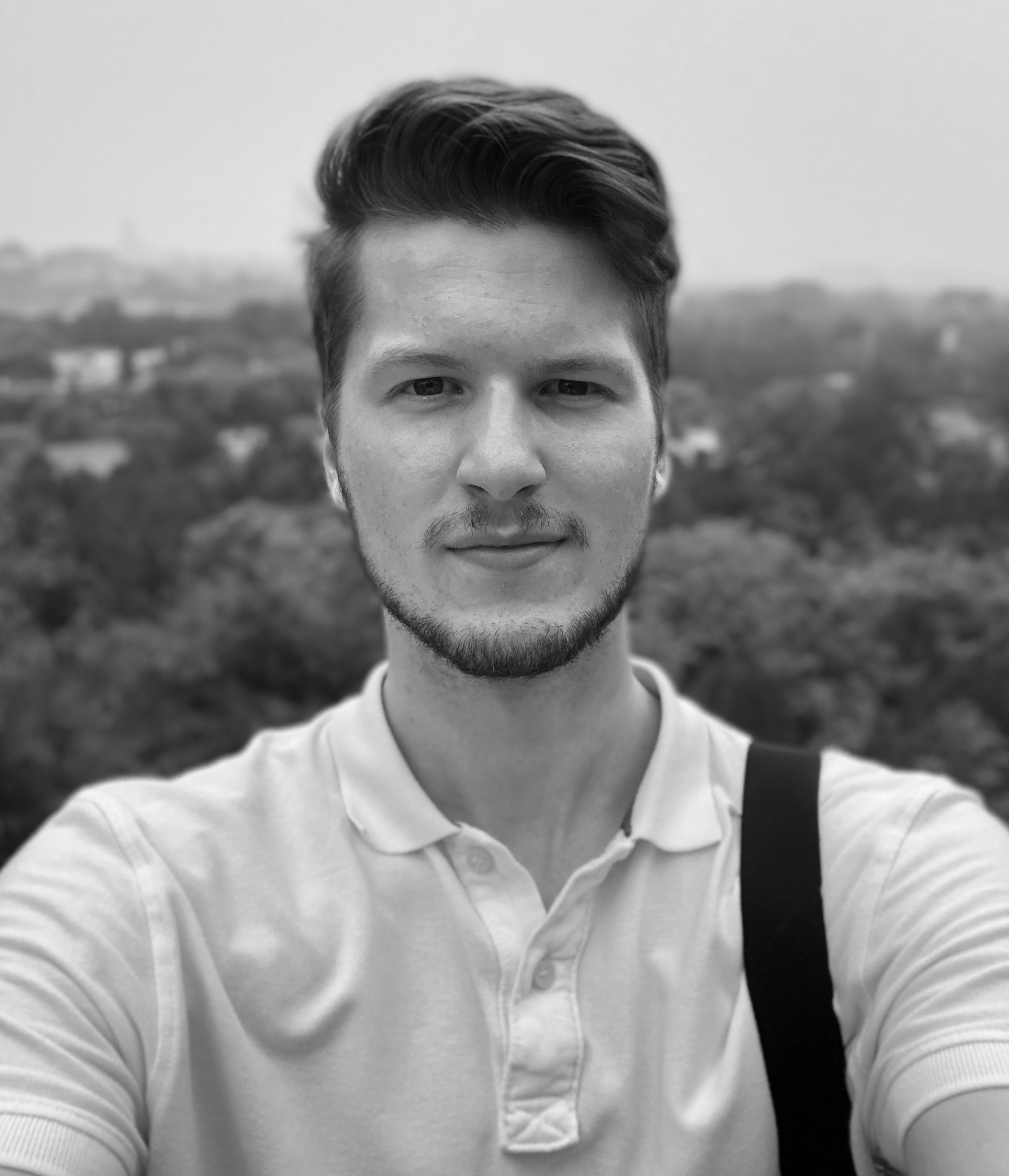}}]
{Martin H. Nielsen}  was born in Aalborg, Denmark. He received his B. Eng degree in electronic systems and his M.Eng. degree in Wireless Communications, from Aalborg University, Denmark in 2017 and 2019 respectively. He is currently a Ph.D. fellow at Aalborg University where he is researching AI, RF, and mm-wave systems. 
His research interests include artificial intelligence, circuits, systems for 5G and 6G, and satellite communication. 
\end{IEEEbiography}

\vspace{-1.2cm}
\begin{IEEEbiography}[{\includegraphics[width=1in,height=1.25in,clip,keepaspectratio]{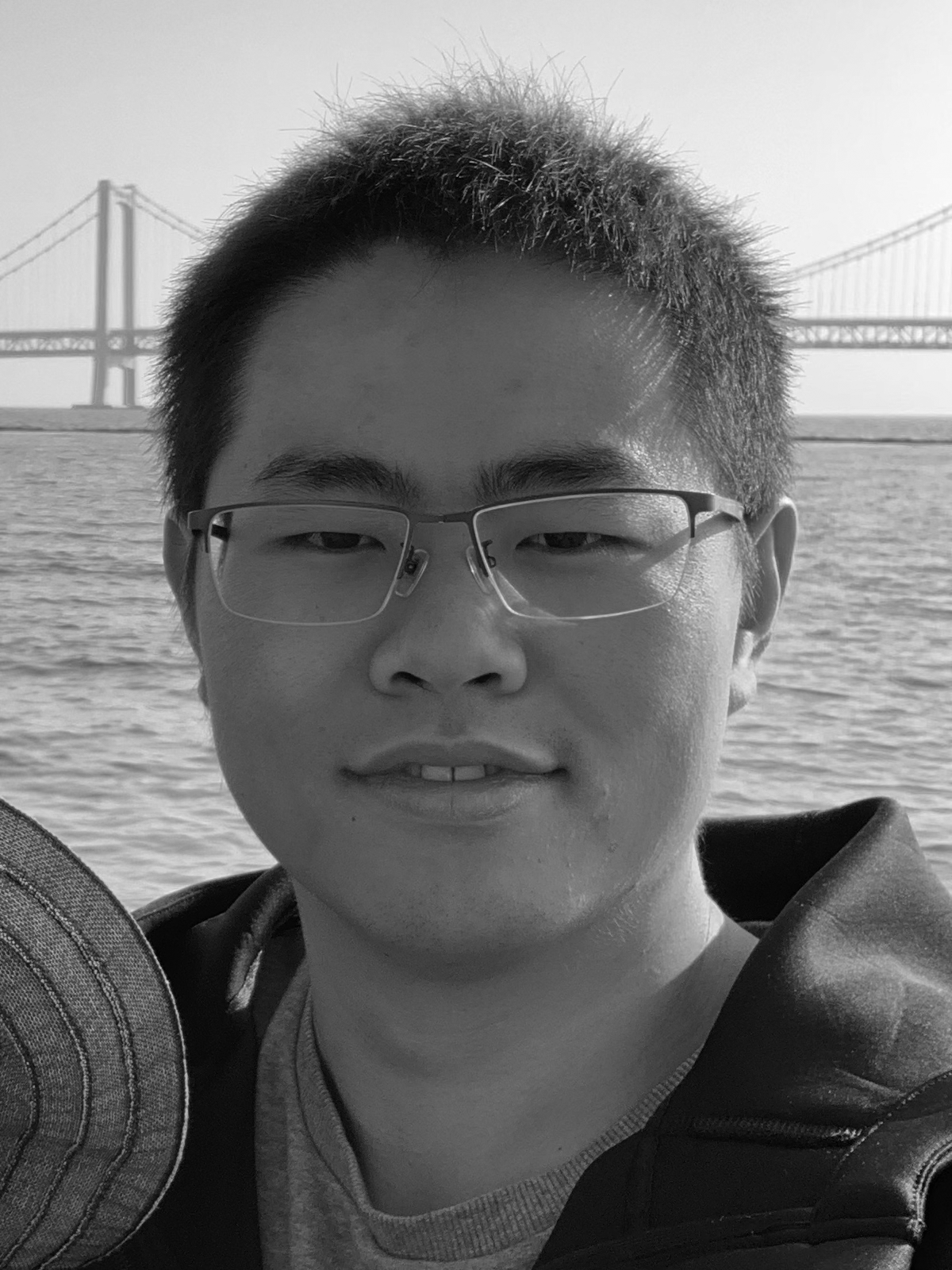}}]
{Yufeng Zhang} was born in Altay, China. He received the B.Eng. degree in electronic engineering from the University of Electronic Science and Technology of China (UESTC) of Chengdu, China in 2015. He is currently working towards the Ph.D. degree in electromagnetic field and microwave technology at the University of Chinese Academy of Sciences (UCAS), Beijing, China. His research interests include artificial neural networks and its application for communication systems. 


\end{IEEEbiography}
\vspace{-1.2cm}
\begin{IEEEbiography}[{\includegraphics[width=1in,height=1.25in,clip,keepaspectratio]{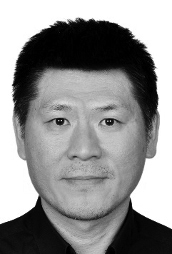}}]
{Changbin Xue} was born in Jinzhou, China. He received his MS. degree in  Electronic Engineering from Harbin Institute of Technology, China, in 1997, and PH.D degree in Condensed Matter Physics from  Beijing Institute of Technology, China, in 2017, respectively. 
He is a professor with the National Space Science Center, Chinese Academy of Sciences. He was the general designer of the scientific payload system in China's Chnag'e-4 lunar mission. His current research interests include data processing and analysis and aircraft system engineering technology.

\end{IEEEbiography}

\vspace{-1.2cm}
\begin{IEEEbiography}[{\includegraphics[width=1in,height=1.25in,clip,keepaspectratio]{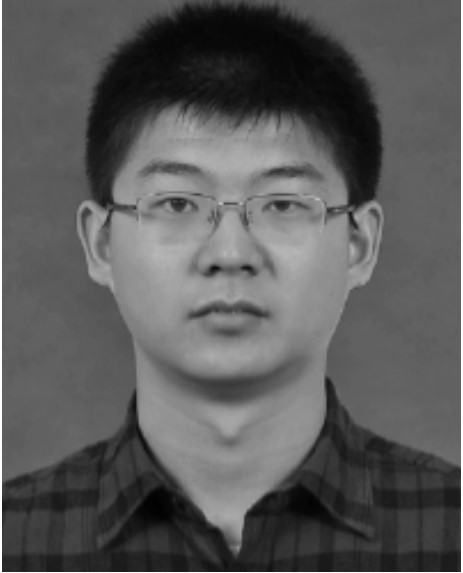}}]
{Jian Ren} was born in Dezhou, China, in 1989. He received the B.Sc. degree in electronic engineering and the M.Eng. degree in electromagnetics and microwave technology from Xidian University, Xi'an, China, in 2012 and 2014, respectively. He received the Ph.D. degree from the City University of Hong Kong, Hong Kong, in 2018. He is currently an Associate Professor with the National Key Laboratory of Antennas and Microwave Technology, Xidian University. His research interests include microwave circuits, metamaterials, RFID, and dielectric resonator antennas.
\end{IEEEbiography}

\vspace{-1.2cm}
\begin{IEEEbiography}[{\includegraphics[width=1in,height=1.25in,clip,keepaspectratio]{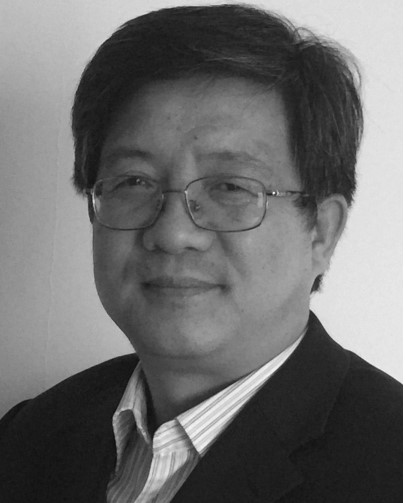}}]
{Yingzeng Yin} received the B.S., M.S., and Ph.D. degrees in electromagnetic wave and microwave technology from Xidian University, Xi'an, China, in 1987, 1990, and 2002, respectively. 
From 1992 to 1996, he was an Associate Professor with the Department of Electromagnetic Engineering, Xidian University, where he has been a Professor since 2004. His current research interests include the design of microstrip antennas, feeds for parabolic reflectors, artificial magnetic conductors, phased array antennas, base-station antennas, and computer aided design for antennas. 
\end{IEEEbiography}

\vspace{-1.2cm}
\begin{IEEEbiography}[{\includegraphics[width=1in,height=1.25in,clip,keepaspectratio]{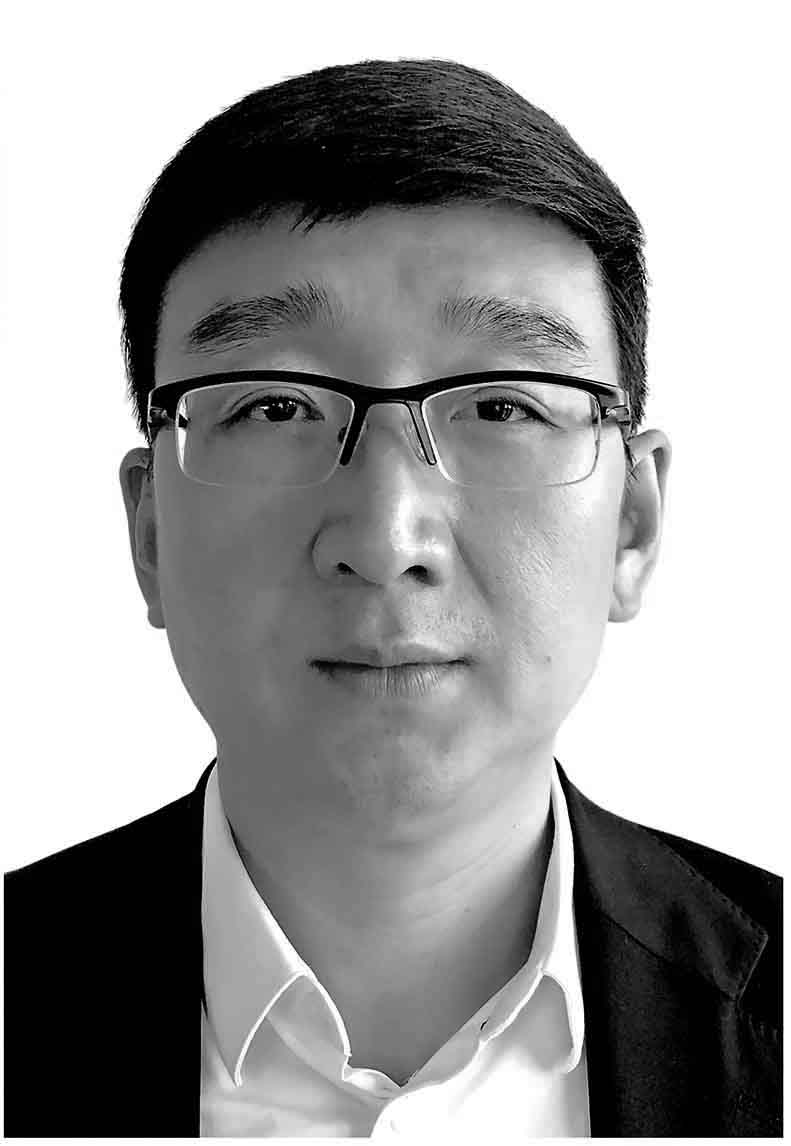}}]
{Ming Shen} (M'11) was born in Yuxi, China. He received his Ph.D. degree in Wireless Communications, with the Spar Nord Annual Best Thesis nomination, from Aalborg University, Denmark, where he is currently an associate professor in RF and mm-wave systems. His research interests include circuits and systems for 5G and satcom, biomedical sensing, and artificial intelligence. He is the grant holder of several Danish national research projects and TPC member of IEEE NORCAS.
\end{IEEEbiography}

\vspace{-1.2cm}
\begin{IEEEbiography}[{\includegraphics[width=1in,height=1.25in,clip,keepaspectratio]{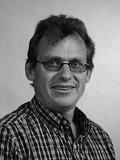}}]
{Gert Fr{\o}lund Pedersen} was born in 1965. He received the B.Sc. and E.E. (Hons.) degrees in electrical engineering from the College of Technology in Dublin, Dublin Institute of Technology, Dublin, Ireland, in 1991, and the M.Sc.E.E. and Ph.D. degrees from Aalborg University, Aalborg, Denmark, in 1993 and 2003, respectively. Since 1993, he has been with Aalborg University where he is a Full Professor heading the Antennas, Propagation and Millimeter-wave Systems LAB. 


\end{IEEEbiography}
\end{document}